\documentclass{PoS}
\usepackage{graphicx}
\usepackage{amsmath}
\usepackage{aasmacros}

\title{Observations of the FSRQ 3C 279 during the flaring state of 2017 and 2018 with H.E.S.S.}

\ShortTitle{3C 279 in 2017/2018 - H.E.S.S.}

\author{\speaker{G.~Emery}$^1$, M.~Cerruti$^{1,2}$, A.~Dmytriiev$^3$, F.~Jankowsky$^4$, H.~Prokoph$^5$, C.~Romoli$^6$, M.~Zacharias$^{7,8}$ for the H.E.S.S. collaboration\footnote{for collaboration list see PoS(ICRC2019)1177}\\
        $^1$Sorbonne Universit\'e, Universit\'e Paris Diderot, Sorbonne Paris Cit\'e, CNRS/IN2P3, Laboratoire de Physique Nucl\'eaire et de Hautes Energies, LPNHE, 4 Place Jussieu, F-75252 Paris, France\\
        $^2$Now at Institut de Ci\`encies del Cosmos (ICCUB), Universitat de Barcelona (IEEC-UB), Mart\'i i Franqu\`es 1, E08028 Barcelona, Spain\\
        $^3$LUTH, Observatoire de Paris, PSL Research University, CNRS, Universit\'e Paris Diderot, 5 Place Jules Janssen, 92190 Meudon, France\\
        $^4$Landessternwarte, Universit\"at Heidelberg, K\"onigstuhl, 69117 Heidelberg, Germany\\
        $^5$DESY, Platanenallee 6, 15738 Zeuthen, Germany\\
        $^6$Max Planck Institut f\"ur Kernphysik, Saupercheckweg 1, 69117 Heidelberg, Germany\\
        $^7$Centre for Space Research, North-West University, Potchefstroom 2520, South Africa\\
        $^8$Theoretische Physik IV, Ruhr-University Bochum, Germany\\
        E-mail: \email{gemery@lpnhe.in2p3.fr}$^*$}
\abstract{The Flat Spectrum Radio Quasar 3C 279 has been very active since a few years with multiple flaring events occurring at high energies. As part of the H.E.S.S. Target of Opportunity program, 3C 279 was observed multiple times in 2017 and 2018 following high states in optical (February and March 2017) or at high energies as seen with \textit{Fermi}-LAT (June 2017, January, February and June 2018). While in January 2018 H.E.S.S. detected an unexpected very high energy (VHE) flare at the end of the MeV-GeV flaring state, in June 2018 it was possible to follow almost continuously the decaying part of a strong \textit{Fermi}-LAT flare, observing with the full array for several nights after the peak of the GeV gamma-ray emission. This has lead to the detection of the source with very high significance. We present here the temporal and spectral results of the H.E.S.S. II dataset together with an overview of the strong multi-wavelength activity seen from 3C 279 between 2017 and 2018.}

\FullConference{36th International Cosmic Ray Conference -ICRC2019-\\
                July 24th - August 1st, 2019 \\
                Madison, Wisconsin, USA}

\begin{document}

\section{Introduction}

The extra-galactic sky at very high energies (VHE; $E>100$~GeV) is dominated by blazars. These sources are active galactic nuclei (AGN) powering a relativistic jet in direction of the Earth. The fluxes at VHE are low enough that individual photons need to be collected over areas of the order of a kilometer square, and strongly variable on time scales ranging from minutes~\cite{2007ApJ...664L..71A} to years~\cite{2017A&A...598A..39H}. The short time variability are particularly interesting for experiments such as H.E.S.S. as daily to weekly increase in fluxes can allow for high quality, time resolved, observations of sources which would otherwise be hard or impossible to see and shorter time variability adds strong constraints on the emission model.

H.E.S.S. is an array of imaging atmospheric Cherenkov telescope (IACT) located in Namibia. The 5 telescopes in the array detect the Cherenkov light from the interaction of a VHE photon in the atmosphere and reconstruct the arrival direction and energy of the original photon. H.E.S.S. can observe in Mono mode using only the largest telescope called CT5 with a mirror diameter of 28 meters, or in Stereo mode using also the smaller CT1-4 with a mirror diameter of 12 meters.

3C 279 is a Flat Spectrum Radio Quasar (FSRQ) located at a redshift of $z=0.536$ which is known for its variability at high energy (HE) and was previously detected by H.E.S.S. during a period of high activity seen with \textit{Fermi}-LAT~\cite{2019arXiv190604996H}. The source was active again in 2017 and 2018 and the observation and analysis of this activity is described in the following sections.

\section{H.E.S.S. Observations}

The AGN Target of Opportunity (ToO) program of H.E.S.S. triggers observations of AGN seen in a high emission state by other experiments and observatories. To do so, public data and privately shared information from optical, X-rays, HE and VHE are checked daily to search for interesting behaviours. In 2017, observations on 3C 279 were triggered twice following high states detected in the optical band. Later in 2017 and in 2018, observations were triggered four additional times following high states detected at HE with \textit{Fermi}-LAT and brought to attention by alerts issued with \texttt{FLaapLUC}~\cite{2018A&C....22....9L}.

Analysis of the data taken for each flare was performed in the Mono mode using CT5 only
 with two independent calibration and analysis chains. The main analysis uses the Model analysis~\cite{2009APh....32..231D} and the cross-check was performed in the ImPACT framework~\cite{2015arXiv150900794M}.

During the bright optical flare of March 2017, both observations by H.E.S.S. led to no detection in respectively 5.1 hours and 4.1 hours of observation time. The HE flare of June 2017 was also not seen at VHE but was only covered by 0.7 hours of observation time.

In January 2018, bad weather delayed H.E.S.S. observations for more than a week. While most of the HE flare was missed, a VHE flare was observed during the night between January 27 and 28~\cite{2018ATel11239....1N}. During the VHE flare night, a detection at a level of significance of 10.7 sigma was achieved with 1.7 hours of observation time. The full campaign amount for a total of 5.0 hours of observation time. The full light curve is displayed on Fig.\ref{fig2}.

The February 2018 HE flare declined when H.E.S.S observation started. No significant excess was obtained with 4.1 hours of observation time.

A full coverage of the decreasing phase of a HE flare seen by \textit{Fermi}, as well as an extensive post-flare surveillance, was achieved in June 2018. A preliminary analysis of the full data set representing 18.7 hours of observation time reaches a significance of 11.8 sigma. This is diluted by the low state observation. 
Since the observation covers multiple days of low flux state at HE, the data set is divided between periods of time corresponding to the high and low state of the source. Further analysis of the high state from MJD 58271 to MJD 58277 gives a significance of 13.5 sigma in 12.6 hours.
A preliminary H.E.S.S. lightcurve is obtained integrating above 120 GeV and assuming a spectrum represented by a power law with spectral index 3.7. A decreasing trend is visible at very high energy as well as at high energy and in the optical (Fig.\ref{fig3}).

\section{Contemporaneous multi-wavelength observations}
\begin{figure*}
  \includegraphics[width=1\textwidth]{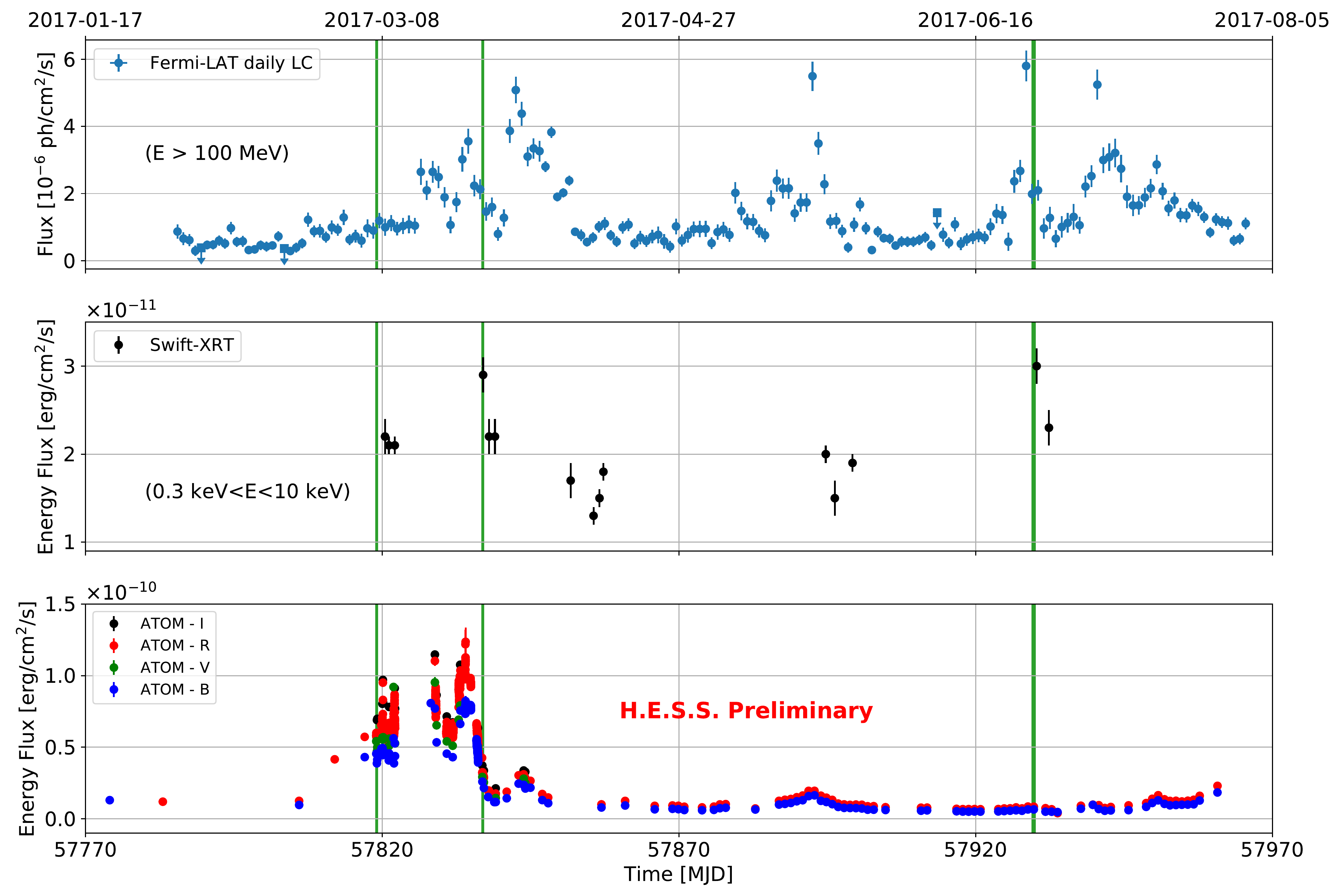}
  \caption{Multi-wavelength observations of 2017. H.E.S.S. time of observation are displayed as green vertical lines.}
\label{fig1}
\end{figure*}
\begin{figure*}
  \includegraphics[width=1\textwidth]{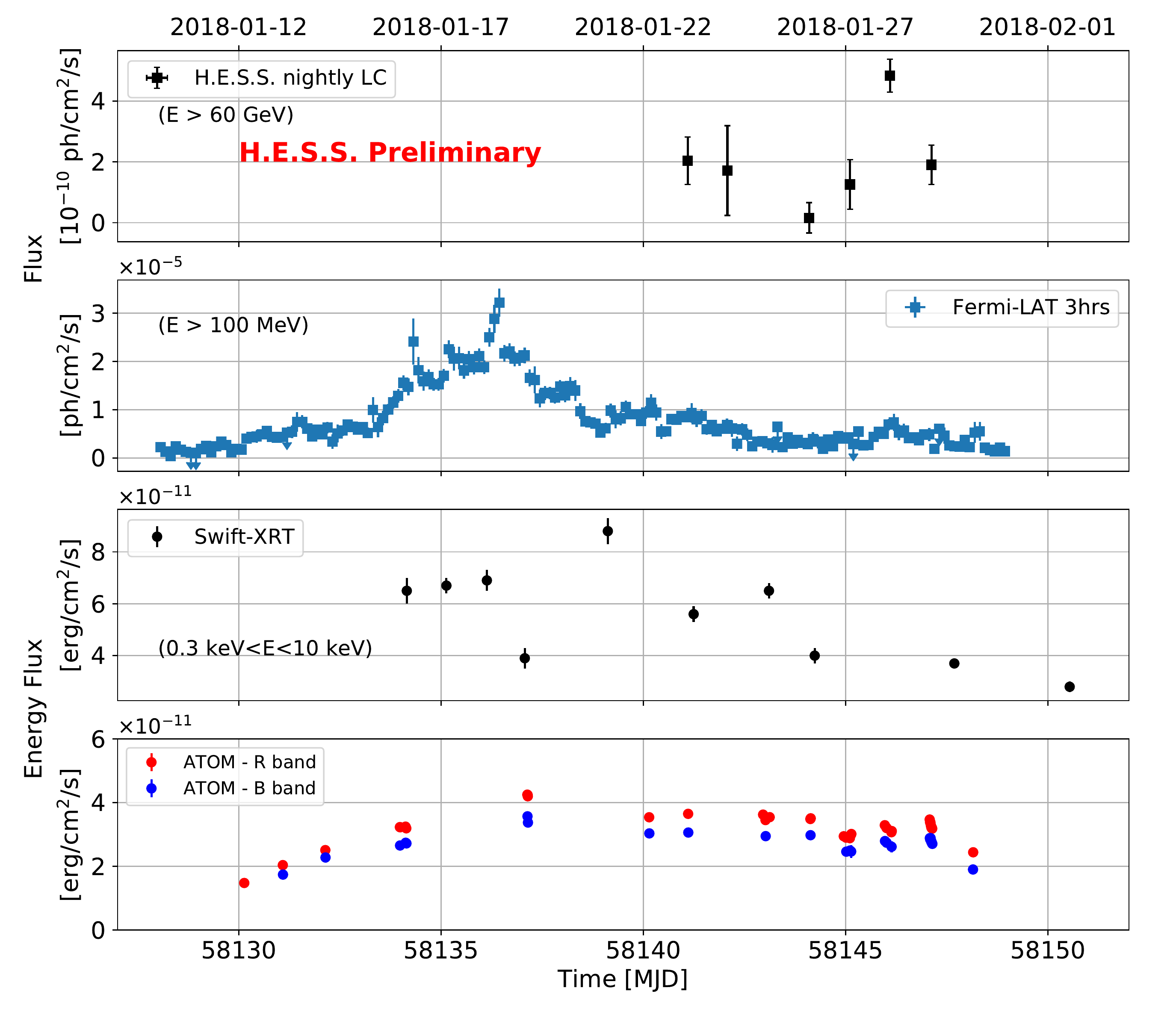}
  \caption{Multi-wavelength observations of January 18. Very different variability behaviour is visible between all wavelengths. A nightly binning is applied on the ATOM observations.}
\label{fig2}
\end{figure*}
\begin{figure*}
  \includegraphics[width=1\textwidth]{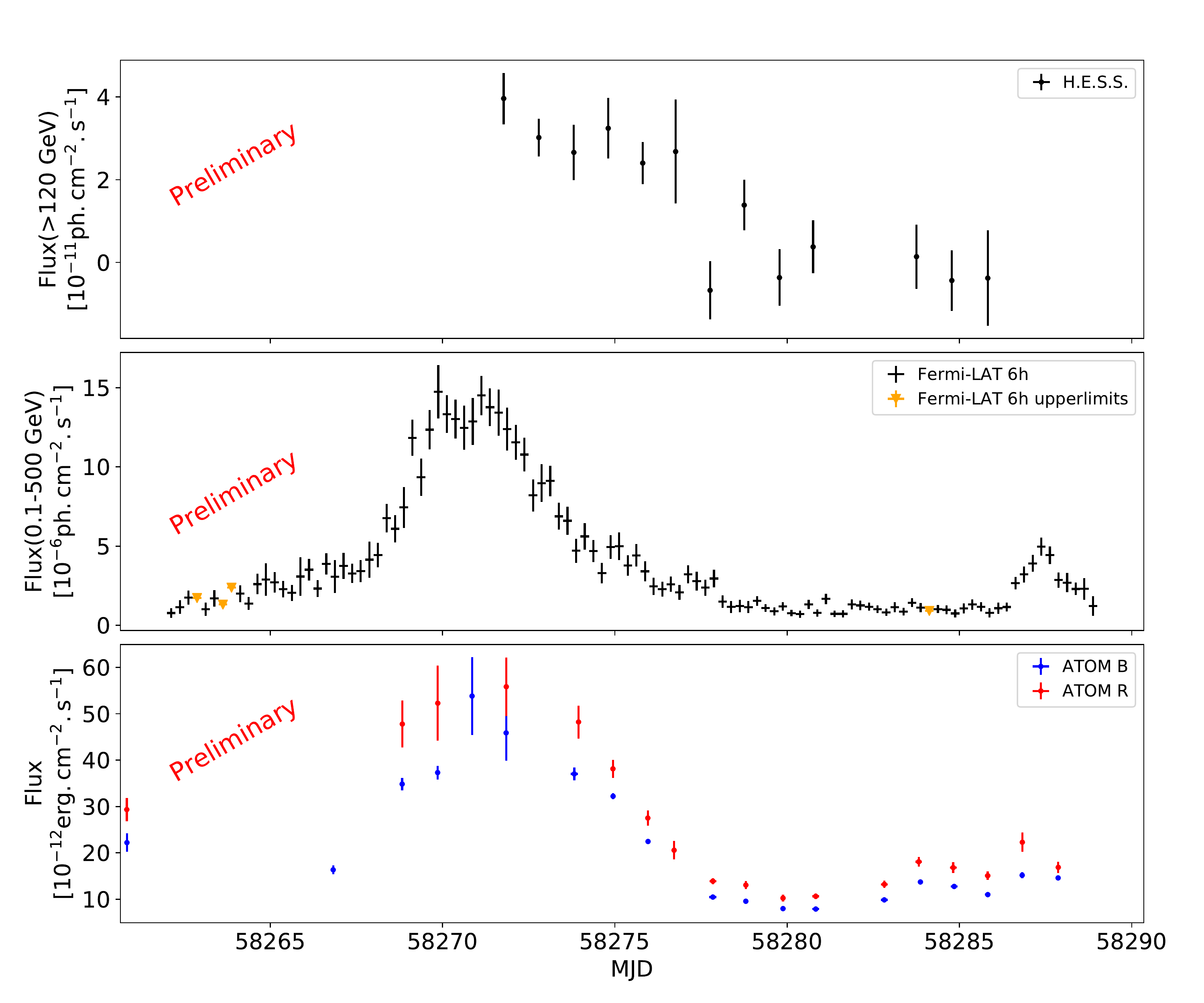}
  \caption{Top panel : H.E.S.S. preliminary light curve above 120 GeV and assuming a powerlaw spectrum with index 3.7. Middle panel : The \textit{Fermi}-LAT light curve with 6 hour bin and flux integrated between 100 MeV and 500 GeV is displayed with black points and orange triangles for the upperlimits. Bottom panel : Optical flux seen by ATOM in the R and B band with a nightly binning}
\label{fig3}
\end{figure*}
\subsection{\textit{Fermi}-LAT data}
The GeV energy range is monitored by the \textit{Fermi}-LAT telescope. The analysis of the data was done using the standard science tools provided by the \textit{Fermi} collaboration (version 5.10.5) with Instrument Response Functions (P8R2) for photons of the SOURCE class (evclass 128, evtype 3) in the energy range 100 MeV - 500 GeV. The region of interest was a square with a side of 20 degrees centred in the source position. The background model was based on the 3FGL catalogue \cite{2015ApJS..218...23A} and the diffuse emission models {\tt gll\_iem\_v06.fits} and {\tt iso\_P8R2\_SOURCE\_V6\_v06.txt}. The final parameters were obtained after a BINNED maximum likelihood fit. The daily lightcurves were obtained from the full time interval models fixing all the background sources except those flagged as variable in the 3FGL. For lightcurves with a shorter time binning, the starting point was the model of the daily interval with all parameters fixed except those of 3C 279.
The data for the 2017 were analysed in the time interval from 2017-02-01 to 2017-08-01 and the lightcurve is shown in figure (Fig.\ref{fig1}). The 2018 dataset was instead analysed from 2018-01-01 to 2018-06-20. Part of the lightcurve are shown on Fig.\ref{fig2} for January 2018 and Fig.\ref{fig3} for June 2018. The data show strong variability of the flux and of the photon index of the source.

\subsection{Swift data}

Following the detection of flaring activity at various wavelenghts, ToO observations have been requested to the Neil Gehrels \textit{Swift} Observatory~\cite{2004ApJ...611.1005G}. \textit{Swift} observed 3C\ 279 with two instruments: the X-Ray Telescope (XRT) and the UV/Optical Telescope (UVOT).
Analysis of Swift observations has been performed using HEASOFT 6.23. All XRT spectra are consistent with a power-law emission absorbed by Galactic material (model tbabs*powerlaw in Xspec). The absorbed integral flux between 0.3 and 10 keV is reported in Figures \ref{fig1} and \ref{fig2}.
Simultaneously with XRT, UVOT observed 3C\ 279 in the UV/optical band. The available data will be described and used in the upcoming publication.

\subsection{ATOM data}

The Automatic Telescope for Optical Monitoring is an optical telescope located at the H.E.S.S. site in Namibia. It provides optical monitoring on known gamma-ray emitters as well as multi-wavelength support for target-of-opportunity events and covered all presented events in R and B bands. Data was reduced and analysed using ATOM's automatic analysis pipeline ADRAS version 2.6.12. Fluxes are obtained via differential photometry using between 2 and 4 custom calibrated comparison stars. Lightcurves are displayed on Fig.\ref{fig1} for 2017, Fig.\ref{fig2} for January 2018 and Fig.\ref{fig3} for June 2018.

\begin{figure*}
\hspace{-0.01\textwidth}
\begin{minipage}[t]{0.5\textwidth}
  \includegraphics[width=1\textwidth]{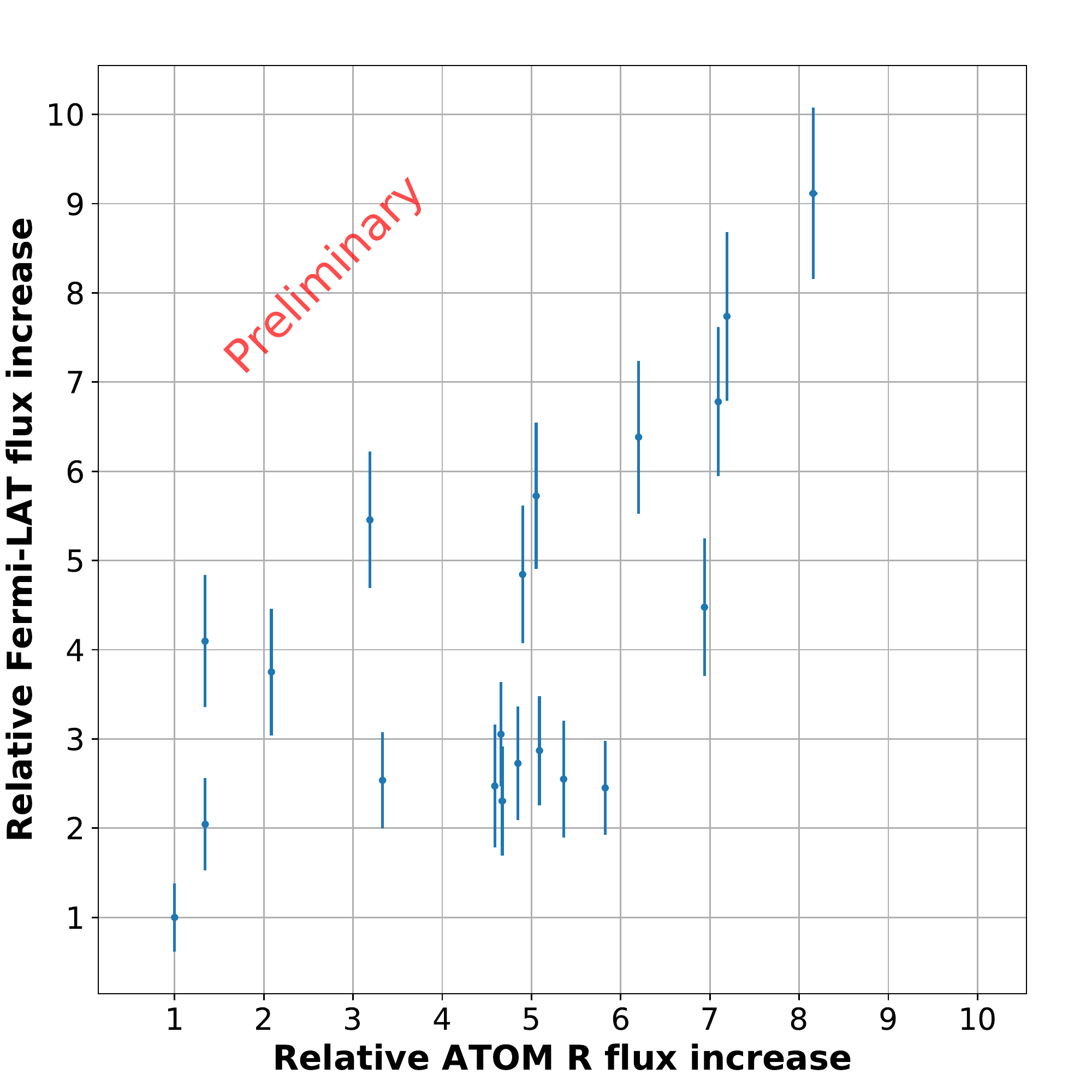}
  \caption{Comparison between the flux variation of \textit{Fermi}-LAT and ATOM R during the optical flare of 2017 from MJD 57805 to 57840}
\label{fig4}
 \end{minipage}
\hspace{0.02\textwidth}
\begin{minipage}[t]{0.5\textwidth}
  \includegraphics[width=1\textwidth]{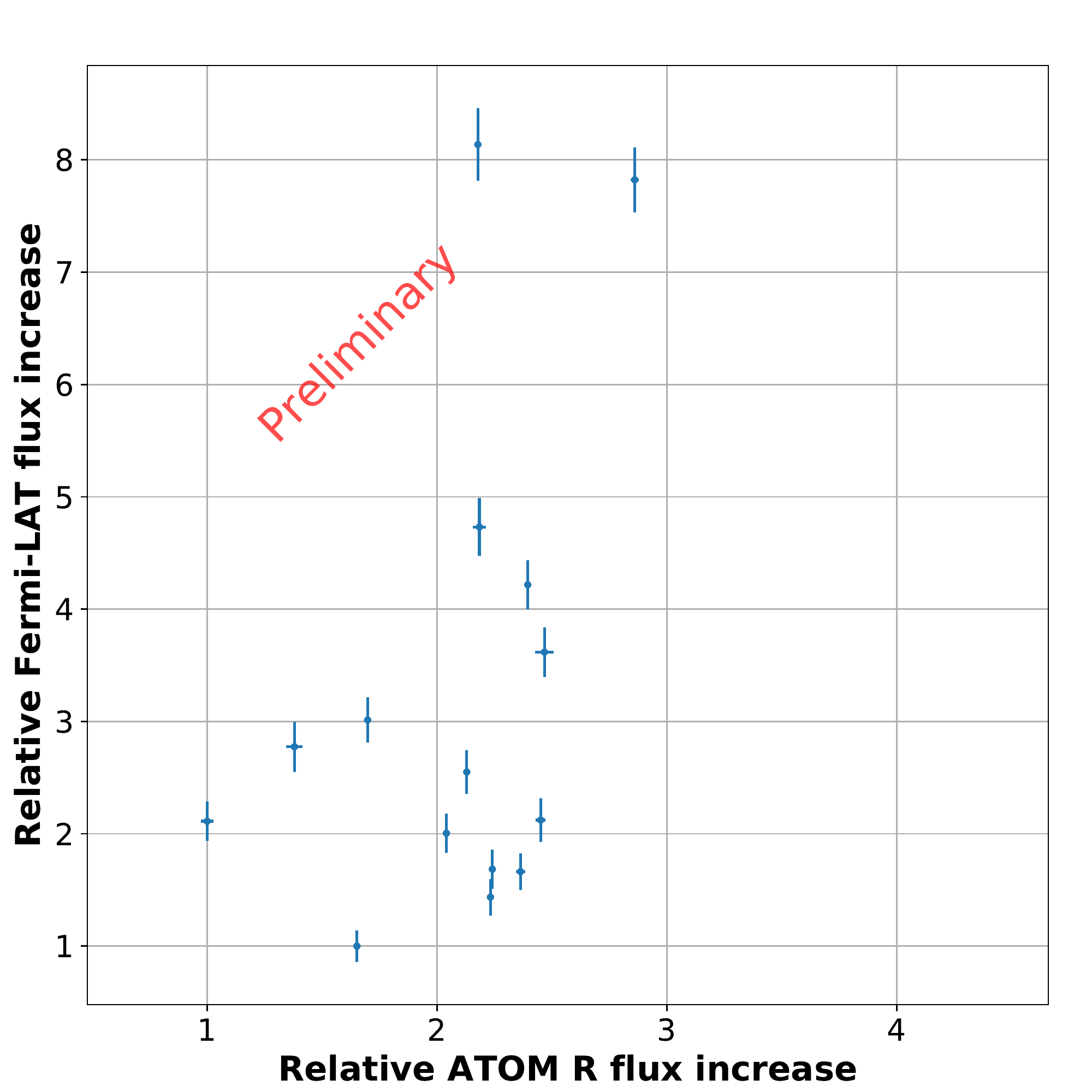}
  \caption{Comparison between the flux variation of \textit{Fermi}-LAT and ATOM R during the January 2018 flare from MJD 58130 to 58148}
\label{fig5}
 \end{minipage}
\end{figure*}

\section{Discussion / Correlations }

The 2017 and 2018 ToO observation campaigns on 3C 279 led to two particularly interesting H.E.S.S. datasets. For the  January 2018 short VHE flare, the high significance 
short time scale and strong VHE variation mirrored by a weaker HE evolution add new interesting properties
compared to the previous detection. The June 2018 dataset on the other hand is particularly interesting due to the extended observation with H.E.S.S. observations covering multiple days of HE activity. This allowed for the strongest detection of 3C 279 by H.E.S.S. to date and access to evolution on time scale of multiple days.

During all those campaigns, multi-wavelength observation were performed. Two behaviours concerning the correlation between optical and HE can be observed. Some flares were visible in HE while the optical flux remained constant. This can be seen with the HE flare after April 2017 (See Fig.\ref{fig3}). While others displayed strong increase in both energy band. For each one when H.E.S.S. also observed, we looked at the correlation between the optical fluxes seen by ATOM and HE fluxes seen by \textit{Fermi}.

To compute the uncertainties on the correlation coefficients, a set of simulated datasets is built. For each point in the correlation plot, 500 new ones are simulated having a 2D tilted Gaussian distribution with the errorbars as width and the Pearson's coefficient derived from the original points as a correlation factor. The final result is a distribution of correlation coefficients. Here we report the 50th percentile as the reference value and we use the 16th and 84th ones to derive the uncertainties on it.

\begin{figure*}
\begin{minipage}[t]{0.5\textwidth}
 \includegraphics[width=1\textwidth]{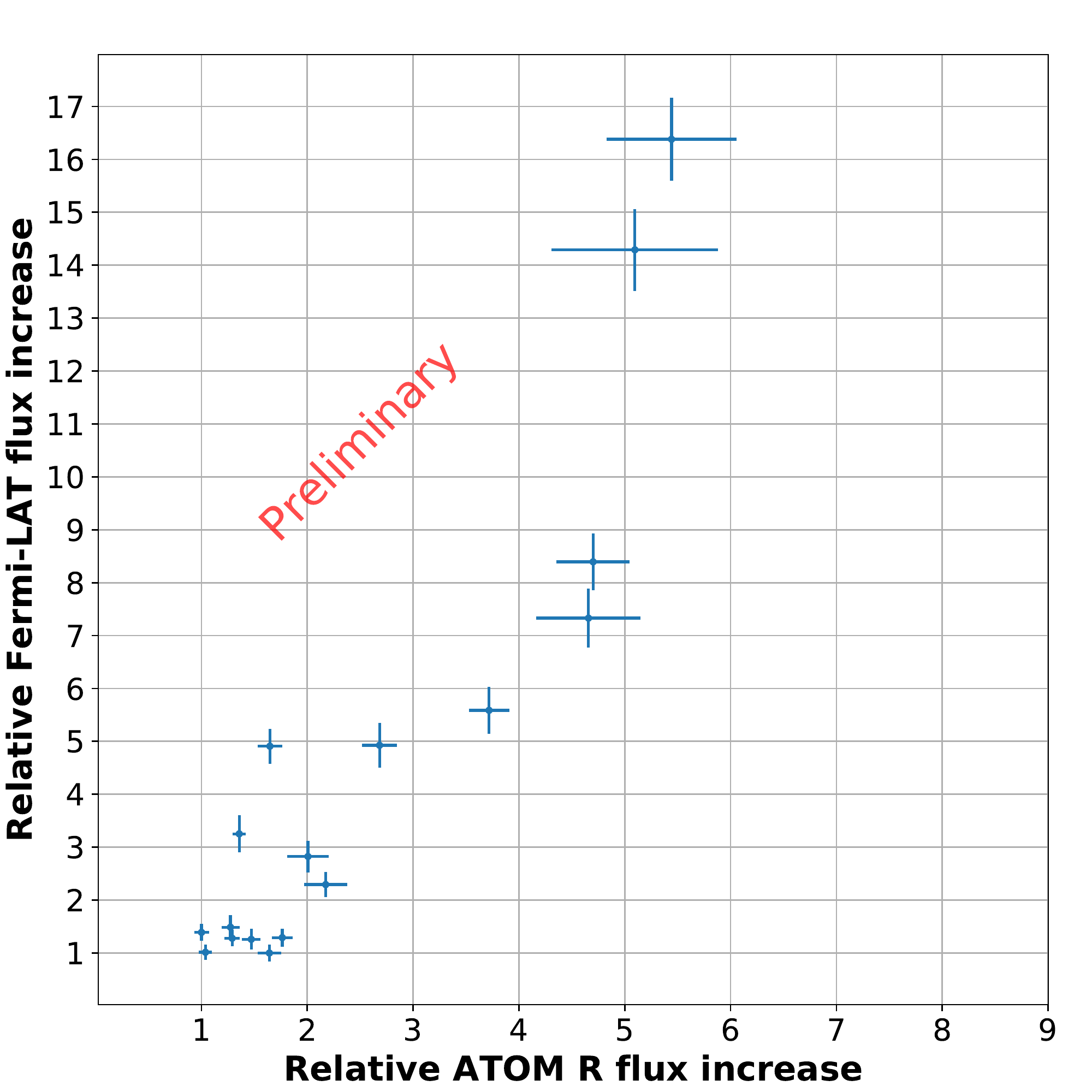}
  \caption{Comparison between the flux variation of \textit{Fermi}-LAT and ATOM R during the June 2018 flare from MJD 58266 to 58288}
\label{fig6}
\end{minipage}
\end{figure*}

Between MJD 57805 and 57840, corresponding to the optically active flares of early 2017, elevated fluxes in optical or HE are associated with elevated fluxes in the other band (See Fig.\ref{fig4}) but the correlation is quite loose (Pearson coefficient = $0.61 \pm 0.05$).
Between MJD 58130 and 58148, corresponding to the January 2018 flare, large HE increase with low optical flux increase as well as optical increase with no HE increase are visible (See Fig.\ref{fig5}). Hence a low correlation (Pearson coefficient = $0.41 \pm 0.02$).
Between MJD 58268 and 58288, corresponding to the June 2018 flare, a strong correlation is visible (See Fig.\ref{fig6}). The Pearson coefficient reaches $0.91 \pm 0.03$ and HE variations appear more than twice as important as optical variations.

\section{Conclusion}

H.E.S.S. detected two particularly interesting flaring behaviour by 3C 279 after triggering on high energy high flux states seen by \textit{Fermi}-LAT. In January 2018, an intense VHE flare was observed days after the end of the HE flare and with a limited simultaneous HE flux increase.
In June 2018, H.E.S.S. achieved a highly significant detection over multiple days of the decreasing part of the HE flare. This flare display a strongly correlated flux evolution between optical and HE and a preliminary H.E.S.S. light curve also hints at a correlation with the very high energy.

\section{Acknowledgements}


\noindent The full H.E.S.S. acknowledgements can be found at the following link :\\ https://www.mpi-hd.mpg.de/hfm/HESS/pages/publications/auxiliary/\\HESS-Acknowledgements-2019.html\\

\noindent We acknowledge the use of public \textit{Fermi}-LAT data.
We warmly thank the Neil Gehrels \textit{Swift} Observatory team for the approval and prompt scheduling of ToO observations.
We acknowledge the contribution of the ATOM collaboration and use of ATOM data.
M.C. has received financial support through the Postdoctoral Junior Leader Fellowship Programme from la Caixa Banking Foundation, grant n.  LCF/BQ/LI18/11630012

\bibliographystyle{JHEP}
\bibliography{bibliography}

\end{document}